\documentclass[twocolumn,prl,aps,preprintnumbers,
superscriptaddress,amsmath,amssymb,amsfonts,nofootinbib,showpacs,floatfix]{revtex4}
\usepackage{ulem}
\usepackage{graphicx,xcolor,psfrag}
\usepackage{afterpage}
\usepackage{mathrsfs}
\newcommand{\bgi}{\begin{itemize}}
\newcommand{\eni}{\end{itemize}}

\newcommand{\bwt}{\begin{widetext}}
\newcommand{\ewt}{\end{widetext}}
\newcommand{\be}{\begin{equation}} \newcommand{\ee}{\end{equation}}
\newcommand{\bea}{\begin{eqnarray}} \newcommand{\eea}{\end{eqnarray}}
\newcommand{\bean}{\begin{eqnarray*}} \newcommand{\eean}{\end{eqnarray*}}

\newcommand{\s}[1]{{\scriptscriptstyle #1}}
\newcommand{\sT}{{\s T}}

\newcommand{\nn}{\nonumber \\}
\def\slash#1{\setbox0=\hbox{$#1$}  % set a box for #1
   \dimen0=\wd0     % and get its size
   \setbox1=\hbox{/} \dimen1=\wd1  % get size of /
   \ifdim\dimen0>\dimen1   % #1 is bigger
      \rlap{\hbox to \dimen0{\hfil/\hfil}} % so center / in box
      #1     % and print #1
   \else     % / is bigger
      \rlap{\hbox to \dimen1{\hfil$#1$\hfil}} % so center #1
      /      % and print /
   \fi}      %   

\begin{document}

\title{A model independent analysis of gluonic pole matrix elements and 
universality of TMD fragmentation functions}

\author{L. P. Gamberg}
\email{lpg10@psu.edu}
\affiliation{Division of Science, 
Penn State University-Berks, Reading, Pennsylvania 19610, USA}

\author{A. Mukherjee}
\email{asmita@phy.iitb.ac.in}
\affiliation{Physics Department,
Indian Institute of Technology Bombay, Powai, Mumbai 400076,
India}

\author{P.J. Mulders}
\email{mulders@few.vu.nl}
\affiliation{
Department of Physics and Astronomy, VU University\\
NL-1081 HV Amsterdam, the Netherlands}

\begin{abstract}
Gluonic pole matrix elements explain the appearance of single spin 
asymmetries (SSA) in high-energy scattering processes. They involve 
a combination of operators which are odd under time reversal (T-odd). 
Such matrix elements appear in principle both for parton 
distribution functions  and parton fragmentation functions.
We show that for parton fragmentation functions these gluonic pole 
matrix elements vanish 
as a consequence of the analytic structure of scattering amplitudes 
in Quantum Chromodynamics.  This result is important in the study of 
the universality of 
transverse momentum dependent (TMD) fragmentation functions.
\end{abstract}

\date{\today}


\maketitle
Cross sections for high energy scattering processes are given by
a convolution of partonic cross sections with parton distribution 
functions (PDF) and parton fragmentation functions (PFF). These functions 
interpreted as momentum distributions and parton decay functions, 
respectively, are given as matrix elements of quark and gluon 
operators~\cite{Collins:1981uw,Jaffe:1991kp,Mulders:1995dh,Boer:1997nt}. 
Among such observables, gluonic pole or Qiu-Sterman matrix
elements involve a combination of operators which is odd under 
time reversal (T-odd). These matrix elements have been extensively 
studied~\cite{Efremov:1981sh,Efremov:1984ip,Qiu:1991pp,Qiu:1991wg,Qiu:1998ia,Kanazawa:2000hz,Eguchi:2006mc,Koike:2006qv}. 
They explain the appearance  of single spin asymmetries (SSA) in high-energy 
scattering processes.  These correlation functions show up
in combination with calculable hard parts 
that may differ from the partonic cross sections through specific calculable 
factors and signs~\cite{Qiu:1998ia,Bacchetta:2005rm,Bomhof:2006ra}. 
In this paper we use general properties of scattering amplitudes 
in Quantum Chromodynamics (QCD) to study the support properties of these 
parton correlation functions. Specifically, assuming unitarity and 
analyticity properties to hold 
for forward parton-hadron scattering amplitudes, we uncover their singularity 
structure which then determines the corresponding properties 
of the PDFs and PFFs.  Using this analysis for fragmentation functions, we show that  single gluon and multi-gluon pole 
matrix elements vanish in the limit when the momenta of these gluons 
become zero.  Since these multi-gluonic pole matrix elements 
appear in  integrated and weighted transverse momentum dependent (TMD) 
fragmentation functions, as a consequence of them being zero, all leading 
T-odd effects in the matrix elements are all part of
 the final state interactions among the fragmentation remnant and final state hadron~\cite{Collins:1992kk,Boer:1997nt}  rather than from 
T-odd partonic operator combinations.
 Thus, when  transverse momentum dependent (TMD) T-odd fragmentation functions appear in observables like SSA they are convoluted with the
standard partonic cross sections.  The vanishing of gluonic pole matrix 
elements for any number of gluons, provides a general proof of the 
universality of these TMD fragmentation functions~\cite{Collins:2004nx}.

 Spectral studies of the gluonic pole matrix elements 
for fragmentation, specifically for quark-quark-gluon matrix elements with
just one gluon field~\cite{Gamberg:2008yt,Meissner:2008yf} 
 already indicated that they vanish.   
Our arguments presented here are more general and can be
applied to multi-gluonic pole matrix elements. They depend 
only on the analytic structure of scattering amplitudes,  yet 
they do not depend on the details of partonic or hadronic masses, 
and are insensitive to integrations over transverse momentum.

We begin our analysis by considering high-energy scattering processes where 
the structure of hadrons is accounted for using quark and gluon correlators, 
which are Fourier transforms of forward matrix elements of non-local quark 
and gluon operators between hadronic states.
For instance the quark-quark correlator 
\bea\label{TMDDF}
\Phi_{ij}^{[\mathcal U]}(x{,}k_\sT)
&=&{\int}\frac{d(\xi{\cdot}P)\,d^2\xi_\sT}{(2\pi)^3}\ e^{ik\cdot\xi}
\nn && \times
\langle P|\,\overline\psi_j(0)\,\mathcal U_{[0;\xi]}\,
\psi_i(\xi)\,|P\rangle\big\rfloor_{\text{LF}}\ ,
\eea
where $\mathcal U_{[\eta;\xi]}
=\mathcal P{\exp}\big[{-}ig{\int_C}\,ds{\cdot}A^a(s)\,t^a\,\big]$
is the gauge link that ensures gauge 
invariance ~\cite{Efremov:1978xm,Boer:1999si,Belitsky:2002sm,Boer:2003cm}.  
The non-locality of the matrix elements needed to describe the 
distribution functions is restricted to the light-front (LF) and
it is convenient to use the Sudakov decomposition 
\be\label{sudakov}
k=x\,P+\sigma\,n+k_\sT, 
\ee
in terms of a generic light-like four-vector $n$ 
satisfying $n^2=0$ and $P\cdot n = 1$. In a particular
hard process, its role is played by
other momenta that are hard with respect to the hadron under 
consideration, e.g.\ $n \approx P^\prime/P\cdot P^\prime$. 
We can then also work with light-cone coordinates. Including mass effects 
one would have $n_- = n$ and $n_+ = P - \tfrac{1}{2}\,M^2\,n$;  with
$k^\pm \equiv k\cdot n_\mp$. These are
$k^+ = k\cdot n = x$ and
$k^- = k\cdot P - \tfrac{1}{2}\,xM^2 = \sigma + \tfrac{1}{2}\,xM^2$.
The transverse momentum is orthogonal to $n$ and $P$. In a hard process
the dependence on $k^-$ of a particular correlator is not important and
it is integrated over, leaving us with the restricted light-front
non-locality $\xi^+ = 0$ (LF). The expansion of these correlators (in Dirac
space) contains the TMD distribution
functions depending on momentum fraction $x$ and transverse momentum
$k_\sT^2$. Upon integration over $k_\sT$ one obtains the collinear
correlators
\bea
\label{colcorrelator}
\Phi(x)= 
%\int d^2k_\sT\ \Phi^{[\mathcal U]}(x{,}k_\sT)
%\nn &=&
{\int}\frac{d(\xi{\cdot}P)}{2\pi}\ e^{i\,x\,\xi\cdot P}
%\nn && \times
\langle P|\,\overline\psi(0)\,\mathcal U_{[0;\xi]}^n\,
\psi(\xi)\,|P\rangle\big\rfloor_{\text{LC}}\,  ,
%\nn
\eea
where non-locality is restricted to the light-cone 
(LC: $\xi\cdot n = \xi_\sT=0$) and the gauge link is
unique, being the straight-line path along $n$. These collinear
correlators are expanded in the 'standard' parton distribution
functions, depending solely on the momentum fraction $x$. 
Here we will not discuss the scale 
dependence~\cite{Collins:1981uk,Ji:2004wu,Collins:2007ph,Bacchetta:2008xw}.

\begin{figure}
\begin{center}
\includegraphics[width=4.5cm]{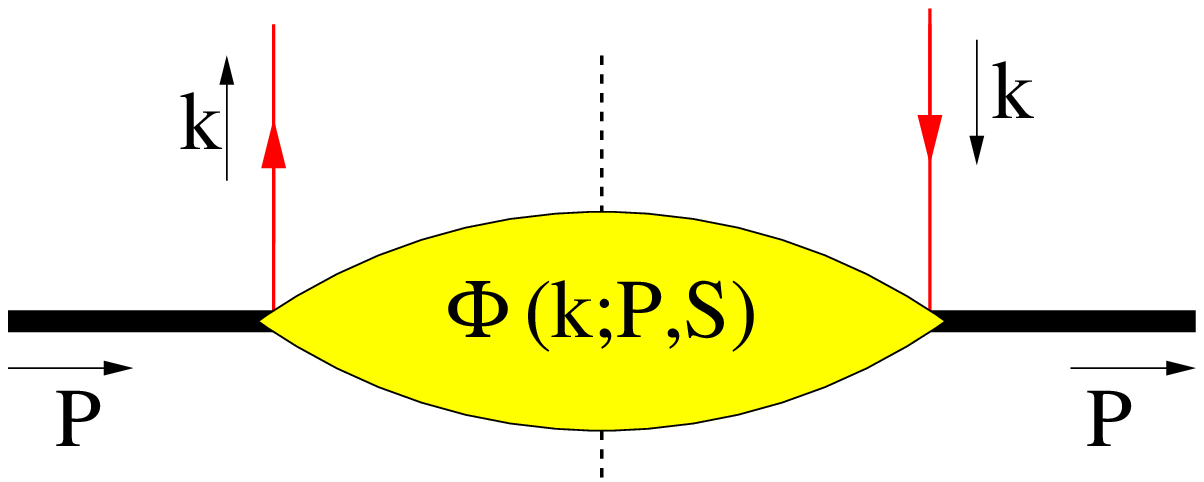}
\hspace{1cm} (a)
\\[0.3cm]
\hspace{1cm}\includegraphics[width=3.4cm]{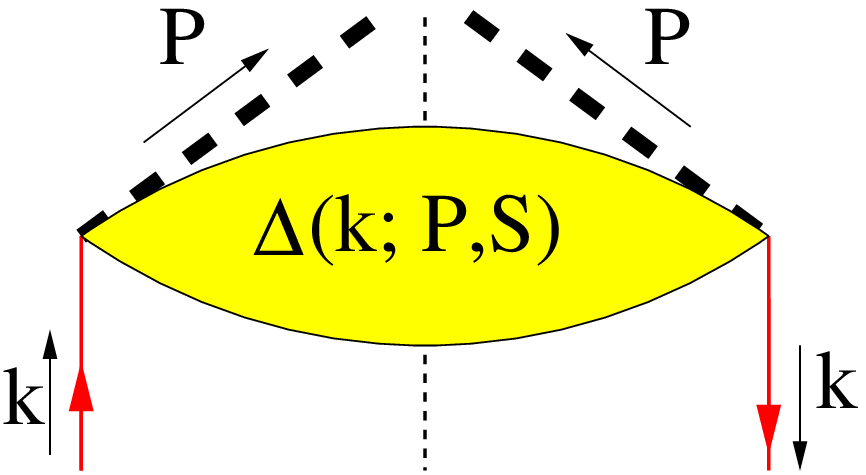}
\hspace{1.3cm} (b)
\\[0.3cm]
\includegraphics[width=4.5cm]{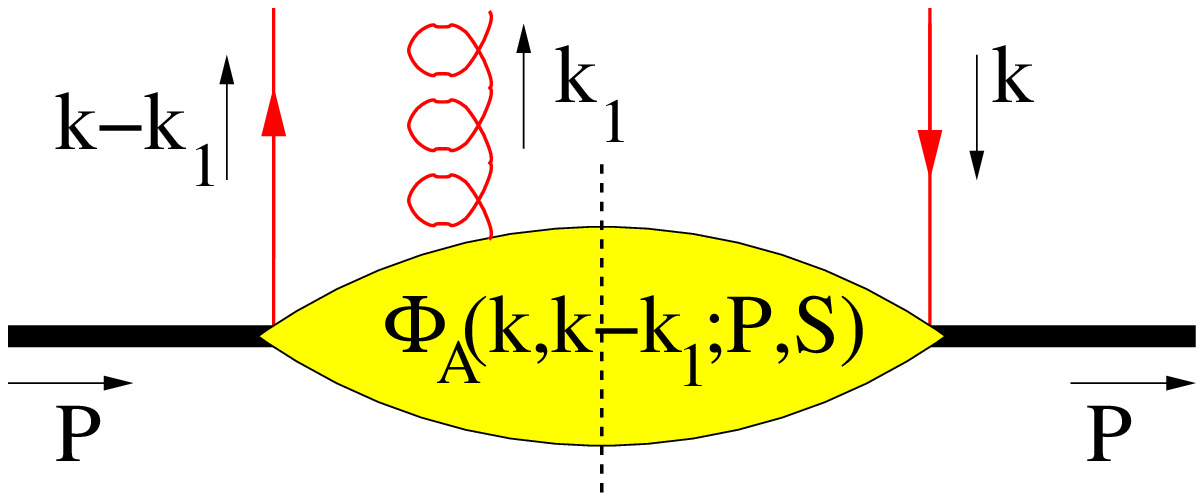}
\hspace{1cm} (c)
\end{center}
\caption{\label{correlators} 
The quark-quark correlators that establish
the {\em non-perturbative} connection between partons and hadrons 
for distribution functions (a) or fragmentation
functions (b) and a quark-quark-gluon 
(multi-parton) correlator (c).} 
\end{figure}

The quark-quark light-front correlator that plays a role in the
fragmentation of partons is
\bea
\Delta^{[\mathcal U]}_{ij}(z,k_\sT) & = &
\sum_X\int\frac{d(\xi\cdot P)\,d^2\xi_\sT}{(2\pi)^3}\ e^{i\,k\cdot\xi}
\langle 0 |\mathcal U_{[0,\xi]}\psi_i(\xi)
\nn && \times
|P,X\rangle \langle P,X|\bar{\psi}_j(0)|0\rangle |_{LF}\,  ,
\label{TMDFF}
%%%\nn
\eea
with the quark momentum, $k = \tfrac{1}{z}\,P + k_\sT + \sigma\,n$,
i.e.\ a Sudakov expansion with $x = 1/z > 1$.
In this case one often refers to the hadron transverse momentum 
$P_\perp = -z\,k_\sT$ (in a frame in which the parton does not have 
a transverse momentum ($k_\perp = 0$)). 
 Diagrammatically these correlators are represented in Fig.~\ref{correlators}.
In this paper, we start the investigation of multi-parton 
correlators  by looking at the case
with one additional gluon, as given in Fig.~\ref{correlators} (c).
They appear in azimuthal asymmetries involving the
$k_\sT$-{\em weighted} correlator,
\bea
\label{TransverseMoment}
\Phi_{\partial}^{\alpha\,[\mathcal U]}(x) 
&=& \int d^2k_\sT\ k_\sT^\alpha\,\Phi^{[\mathcal U]}(x{,}k_\sT)
\nn
&=& \tilde \Phi_\partial^{\alpha}(x)
+C_G^{[\mathcal U]}\,\pi\Phi_G^{\alpha}(x,x)\, ,
\label{DECOMPOSITION}
\eea
which is decomposed in pieces $\tilde \Phi_\partial$ and $\Phi_G$,
that contain T-even and T-odd operator combinations, 
respectively~\cite{Boer:2003cm}. The T-odd parts come with calculable gluonic 
pole factors $C_G^{[\mathcal U]}$ that depend on the gauge link. 
In this paper we focus on the connection of this part to
the zero momentum ($x_1\rightarrow 0$)
limit of a quark-quark-gluon correlator
\bea
&&\Phi_G^\alpha(x,x{-}x_1) =  
\int\frac{d(\xi{\cdot}P)}{2\pi}\frac{d(\eta{\cdot}P)}{2\pi}\ 
e^{ix_1(\eta\cdot P)}e^{i(x-x_1)(\xi\cdot P)}\,
\nn &&\mbox{}\hspace{1cm}\times
\langle P|\,\overline\psi(0)\,U_{[0;\eta]}^n\,gG^{n\alpha}(\eta)
\,U_{[\eta;\xi]}^n\,\psi(\xi)\,|P\rangle\,\big\rfloor_{\text{LC}},
\label{GP}
\eea
where $G^{n\alpha} = n_\mu\,G^{\mu\alpha}$ represent specific components 
of the color field strength tensor ($\alpha$ being transverse). 
The zero momentum limit of this correlator is the gluonic pole
matrix element mentioned above. It is the support of $\Phi_G(x,x-x_1)$ that
we are after and specifically the model independent proof
that it vanishes in the limit $x_1\rightarrow 0$
for the case of the fragmentation correlators ($\vert x\vert > 1$).

\begin{figure}[t]
\begin{center}
\includegraphics[width=8.5cm]{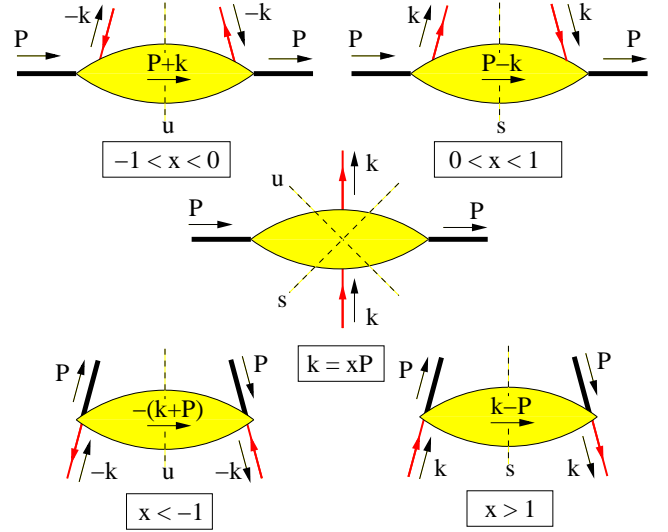}
\end{center}
\caption{\label{interpretation} Integrating parton correlators over $k^-$ 
allows connecting them to a single anti-parton - hadron scattering four-point
function ${\mathcal A}(k^2;s,u)$ (middle). Depending on the value of $x$, 
the imaginary part of this amplitude represents the (anti)-parton 
distribution or fragmentation correlators.}
\end{figure}
The first step in all considerations is the observation that the $k^-$ and 
$k_1^-$-integrations in the quark-quark and quark-quark-gluon correlators
lead to light-front correlators, for which time-ordering is irrelevant.
Therefore the matrix elements can be considered as forward
matrix elements of time-ordered products of operators. 
These represent scattering amplitudes and their
analytic structure 
enables one to make statements about the support of the associated 
parton correlation functions. This can be done 
for quark-quark~\cite{Landshoff:1971xb} and multi-parton 
correlators~\cite{Jaffe:1983hp} and for TMD as well as 
collinear correlators~\cite{Diehl:1998sm}. 
In this language the diagrams of Figs.~\ref{correlators} are 
just hadron-parton amplitudes, e.g.\ the 
quark-quark correlator   Eq.~(\ref{colcorrelator}) related
to the forward antiquark-hadron scattering amplitude ${\mathcal A}(k^2;s,u)$ 
(see Fig.~\ref{interpretation}). Depending on the
precise structure these are untruncated Green functions (time-ordered)
or related to such Green functions via the 
LSZ formalism~\cite{Itzykson:1980rh}. 

The second step is the study of the analytic structure of an amplitude
and in particular the singularities arising from cuts in the forward 
amplitudes. These are  cuts in the (untruncated) legs, 
in particular the parton virtuality $k^2$, and the Mandelstam invariants. 
 The  virtualities are conventionally placed just below 
the real axis, or the invariants are replaced 
by $p^2 + i\epsilon$, $s+i\epsilon$. The integrations over $k^-$ and $k_1^-$ imply integrations over some 
of the invariants. At this point one must make the standard assumption 
that it is possible to use analyticity for QCD-amplitudes. We illustrate 
this step first for the standard quark-quark correlators. In that
case one works with the four-point Green functions, shown in the middle of 
Fig.~\ref{interpretation}, where the cut amplitude depending on the 
particular value of $x$ gives the quark distributions ($0< x< 1$) or the
quark fragmentation functions ($0 < z < 1$ or $x > 1$) 
while analytic continuation to negative $x$ describes the anti-quark
distributions ($-1 < x < 0$) and anti-quark fragmentation 
($-1 < z < 0$ or $x < -1$). 

\begin{figure}[b]
\begin{center}
\includegraphics[width=6cm]{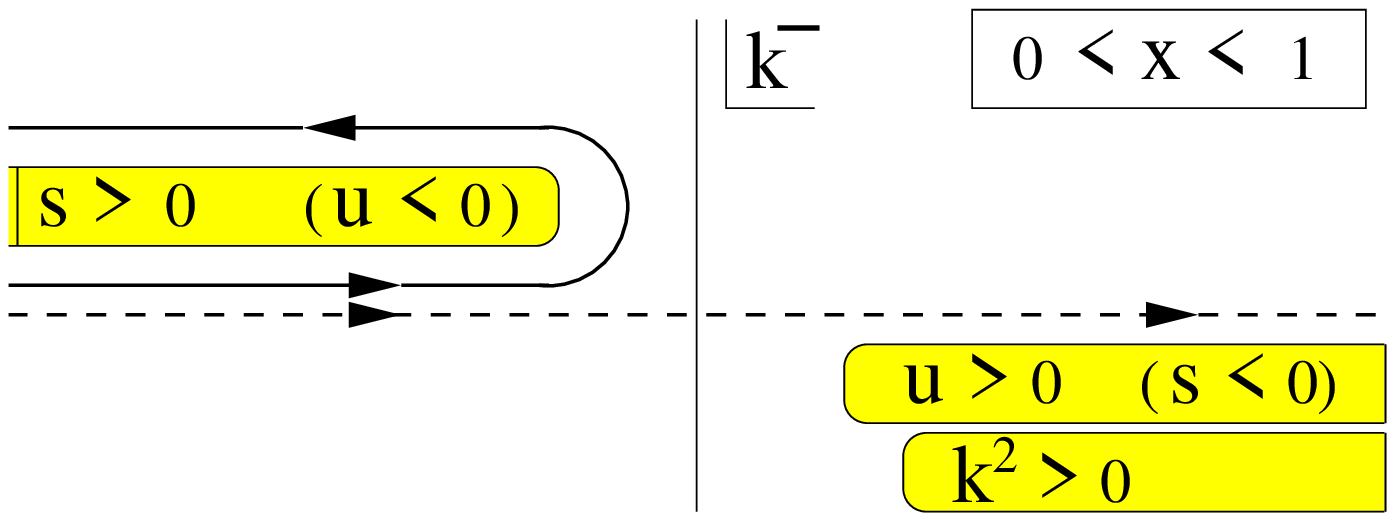}
(a) quarks
\\[0.3cm]
\includegraphics[width=6cm]{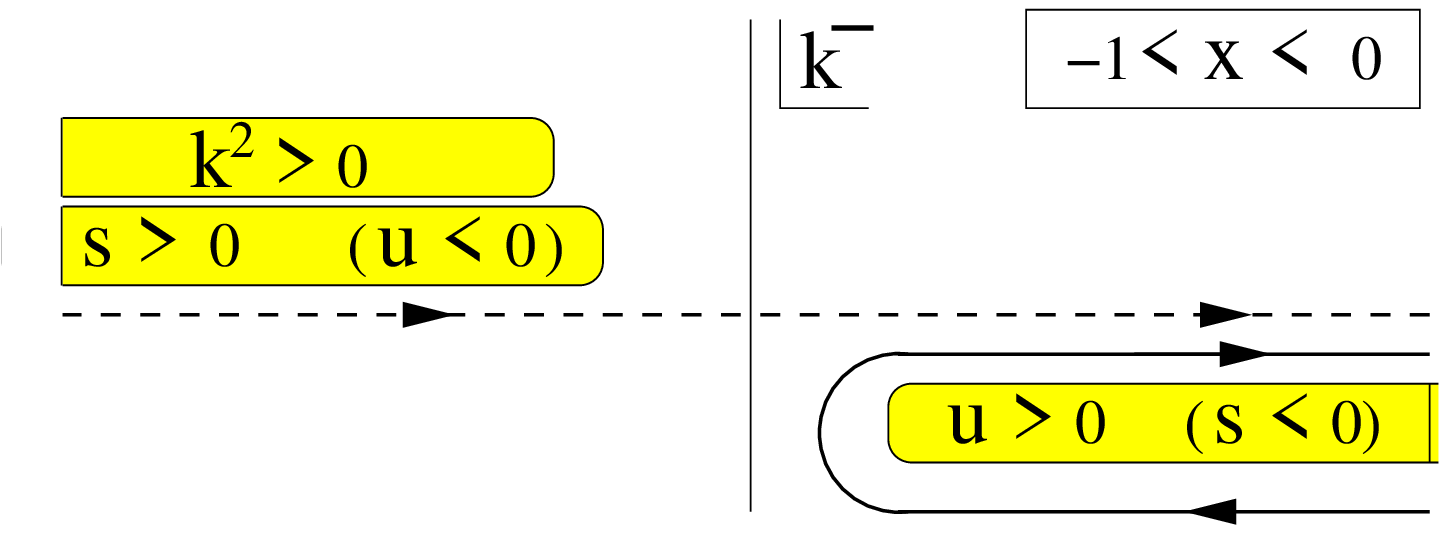}
(b) antiquarks
\end{center}
\caption{\label{singularity-structure} The integration contours
for the $k^-$ integration with respect to the kinematic singularities
in the (forward) anti-parton - hadron scattering amplitude for the case
of (non-vanishing) distribution functions for quarks (a) and
antiquarks (b).}
\end{figure}
The forward amplitude itself has singularities (cuts) for positive 
parton virtuality $k^2$ and in the Mandelstam variables, for which we 
choose $s = (P-k)^2$ and $u = (P+k)^2$. For the forward amplitude,
the invariants are constrained to $s + u = 2\,k^2 + 2\,M^2$, where we
will neglect the hadron masses (they don't play any essential role in
our proof). These singularities then constitute the $s$-cut (for 
$s > 0$ and $u < 0$) and the $u$-cut (for $u > 0$ and $s < 0$).
Using the expansion for $k$ in Eq.~(\ref{sudakov}), one has
$k^2 = 2xk^- + k_\sT^2$ or
$s = 2(x-1)k^- + k_\sT^2$ and a similar expression for $u$.  The 
transverse momenta, just as any of the parton or hadron masses, 
have little bearing on our results so
we omit them in the expressions for $k^-$,
\bea
k^- = \frac{s + i\epsilon}{2(x-1)}
= \frac{u + i\epsilon}{2(x+1)}
= \frac{k^2 + i\epsilon}{2x},
\label{disrel}
\eea
and one sees that for distribution functions the $k^-$ integration with
respect to $s$, $u$ and $k^2$ singularities follows the (dashed) contour in
Fig.~\ref{singularity-structure}. 
The integration contours can be wrapped around the $s$ and $u$-cuts
for positive and negative $x$-values respectively if $\vert x\vert < 1$,
cuts that (in $k_1^-$) smoothly vanish when $\vert x\vert \rightarrow 1$.
Neither masses, nor transverse momenta matter
and the support properties are valid for collinear and TMD PDFs. We get 
\be
\Phi(x) \hspace{-0.02cm}=\hspace{-0.02cm} \theta(x)\,\theta(1-x)\,{\rm Disc}_{[s]}\mathcal A
%\nn && 
+\theta(-x)\,\theta(1+x)\,{\rm Disc}_{[u]}\mathcal A.
%\nn
\ee
As discussed for instance in Ref.~\cite{Polkinghorne:1980mk}, 
the case for fragmentation is
different since one in essence discusses the parton propagator for 
positive $k^2$ (sitting on the cut). For $\vert x\vert > 1$, one simply 
has
\bea
\Delta (x) & = & \theta(x-1)\,{\rm Disc}_{[s]}\mathcal A
+ \theta(-1-x)\,{\rm Disc}_{[u]}\mathcal A.
%\nn &=& \theta(z)\,\theta(1-z)\,{\rm Disc}_{[s]}\mathcal A
%+\theta(-z)\,\theta(1+z)\,{\rm Disc}_{[u]}\mathcal A.
%\nn &&
\eea
In wrapping the integration around the $s$- or $u$-cut we have to assume
convergence in the variable $k^-$ (or $k^2$), or use subtracted relations. 

\begin{figure}[t]
\begin{center}
\includegraphics[width=5.5cm]{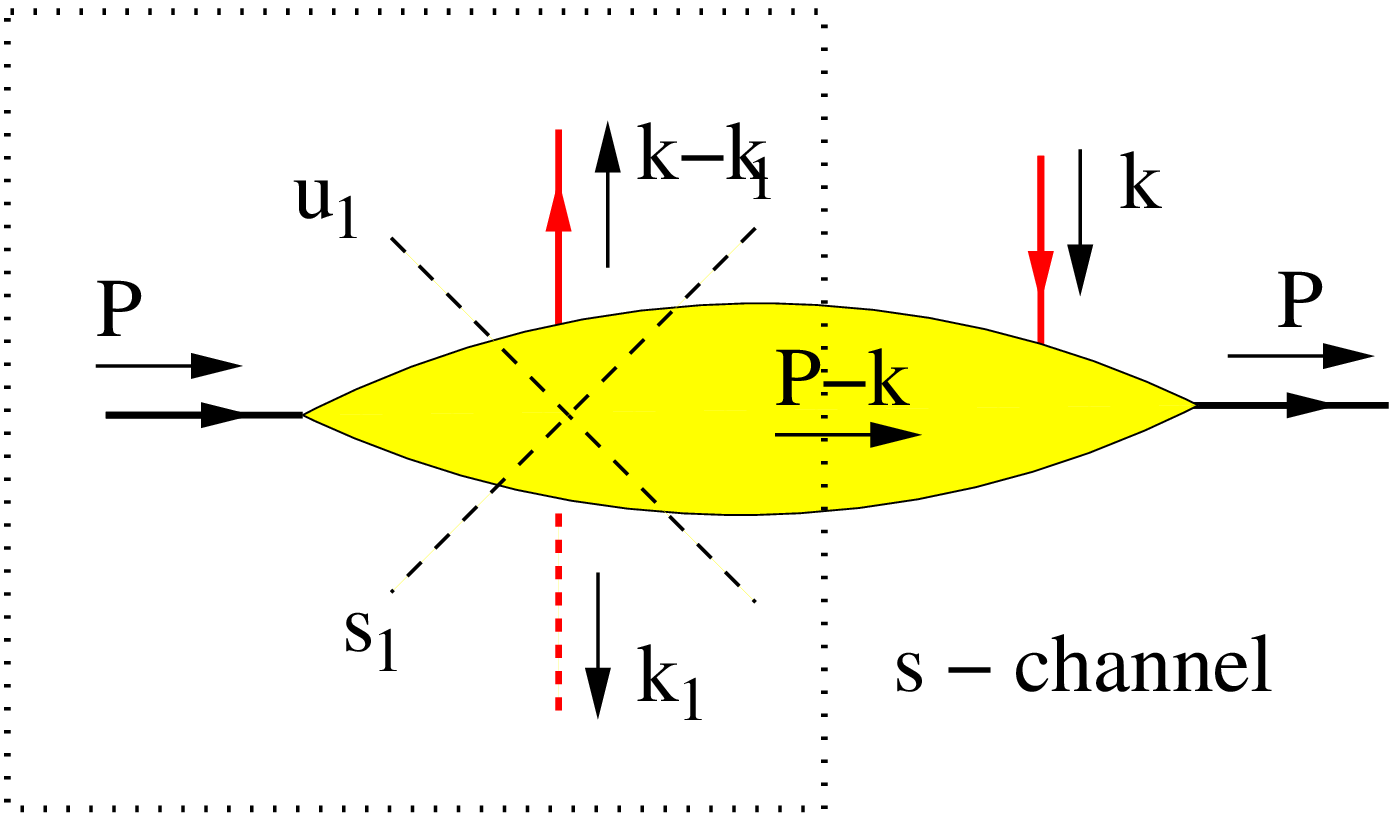}
\hspace{1cm} (a)
\\
\includegraphics[width=5.5cm]{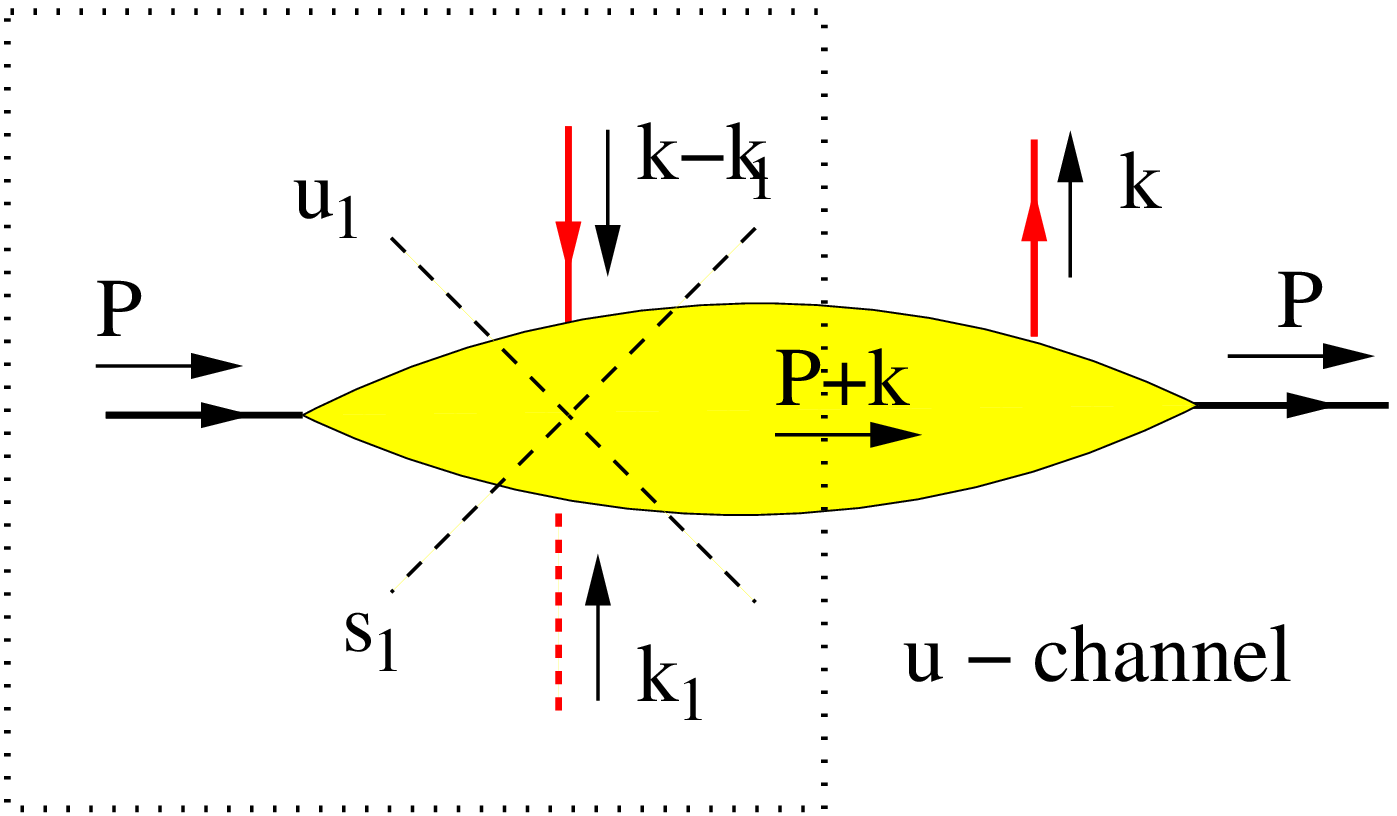}
\hspace{1cm} (b)
\end{center}
\caption{\label{gluonicpole-cuts} The additional invariants for the 
amplitude ${\mathcal A}(k^2;s,u;s_1,u_1;k_1^2,(k-k_1)^2)$ 
relevant for gluonic pole matrix elements, (a) for the 
case $s > 0$ and (b) for the case $u > 0$.
}
\end{figure}
It is important to mention here that the integration of 
Eq.~(\ref{TMDDF}) 
and ${\mathcal A}(k^2;s,u)$  over $k_{\sT}$
leads to ultraviolet 
divergences~\cite{Jaffe:1983hp,Diehl:1998sm,Bacchetta:2008xw}. 
However, this does not invalidate the assumption that the
integral over $k^-$ alone is sufficiently well behaved 
when $k^+$ and $k_\sT$ are fixed. 
Integration over $k_{\sT}$ as well as weighting with $k_\sT$
is anyway intimately linked with QCD evolution of TMDs~\cite{Aybat:2011zv}.

We can extend this analyticity  analysis to the multi-parton distribution and
fragmentation functions in  Eq.~(\ref{GP}), 
by looking at the multi-parton amplitude 
${\mathcal A}(k^2;s,u;s_1,u_1;k_1^2,(k-k_1)^2)$ 
shown in Fig.~\ref{gluonicpole-cuts},
by studying the contours for the additional integrations. 
Defining the momentum of the additional parton as $k_1$ as shown in Fig.~\ref{correlators}, 
one retains the definitions and relations for $s$ and $u$. 
For given positive $s$ ($s$-channel, $x > 0$) or positive $u$ 
($u$-channel, $x < 0$) one 
gets additional invariants $s_1 = (P\mp k\pm k_1)^2$ and $u_1 = (P\mp k_1)^2$
 (cf.\ Figs~\ref{gluonicpole-cuts} (a) and (b), respectively). 
Note that $t_1 = k^2$ in both cases. Furthermore one has parton virtualities.
Depending on if one is dealing with the $s$- or $u$-cut discontinuities,
one has slightly different constraints for $s_1 + u_1$, but for given
values of $s$ and $u$ in the two cases of Fig.~\ref{gluonicpole-cuts}, one 
has cuts along $s_1 > 0$ ($u_1 < 0$) and $u_1 > 0$ ($s_1 < 0$) as well as
for positive parton virtualities.
The relevant singularities for $k_1^-$ are found from 
\bea
k_1^-&=&\frac{s_1+i\epsilon}{2(x_1-(x\mp 1))}+k^-
=\frac{u_1+i\epsilon}{2(x_1\mp 1)}
\nn  &=&
\frac{k_{1}^2+i\epsilon}{2x_1}
=\frac{(k-k_1)^2+i\epsilon}{2(x_1-x)}+k^-
\label{virts}
\eea
(with $\mp$ referring to $s$- and $u$-channel cuts, respectively).
\begin{figure}[t]
\includegraphics[width=6.0cm]{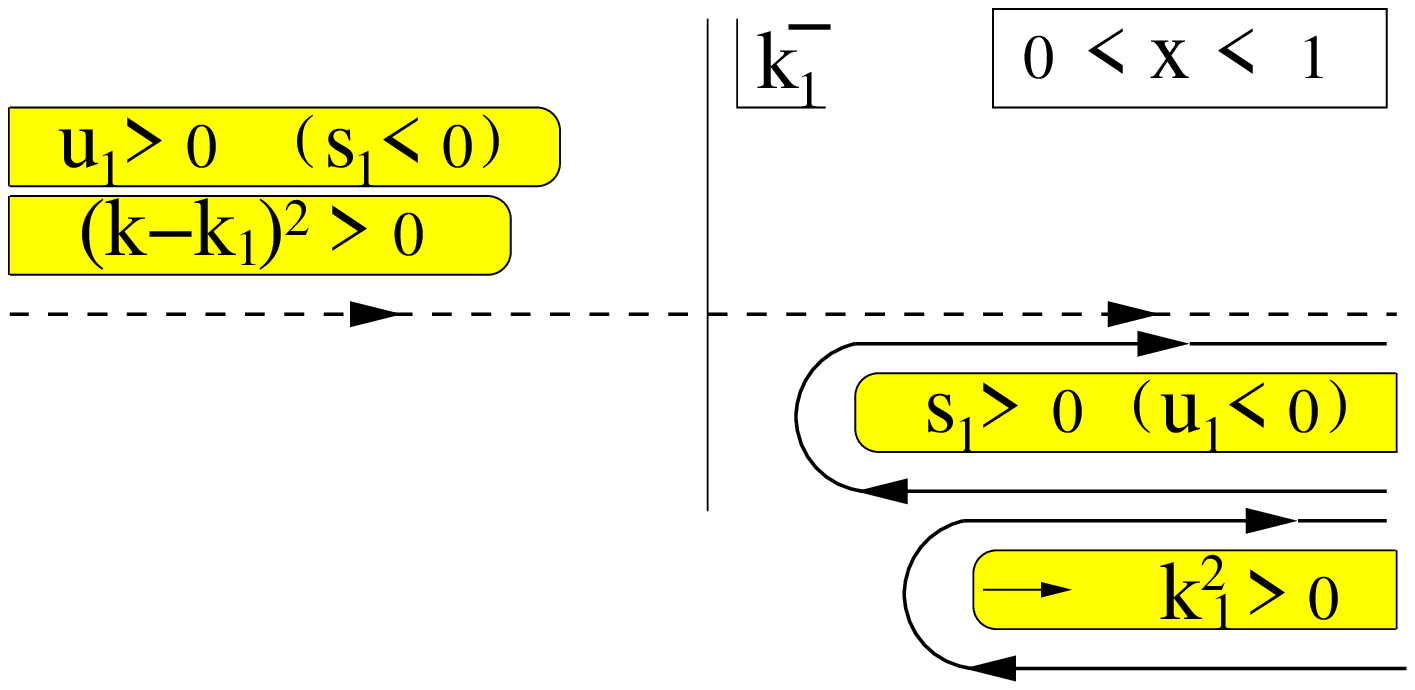}
(a) $\Phi_G(x,x)$
\\[0.3cm]
\includegraphics[width=6.0cm]{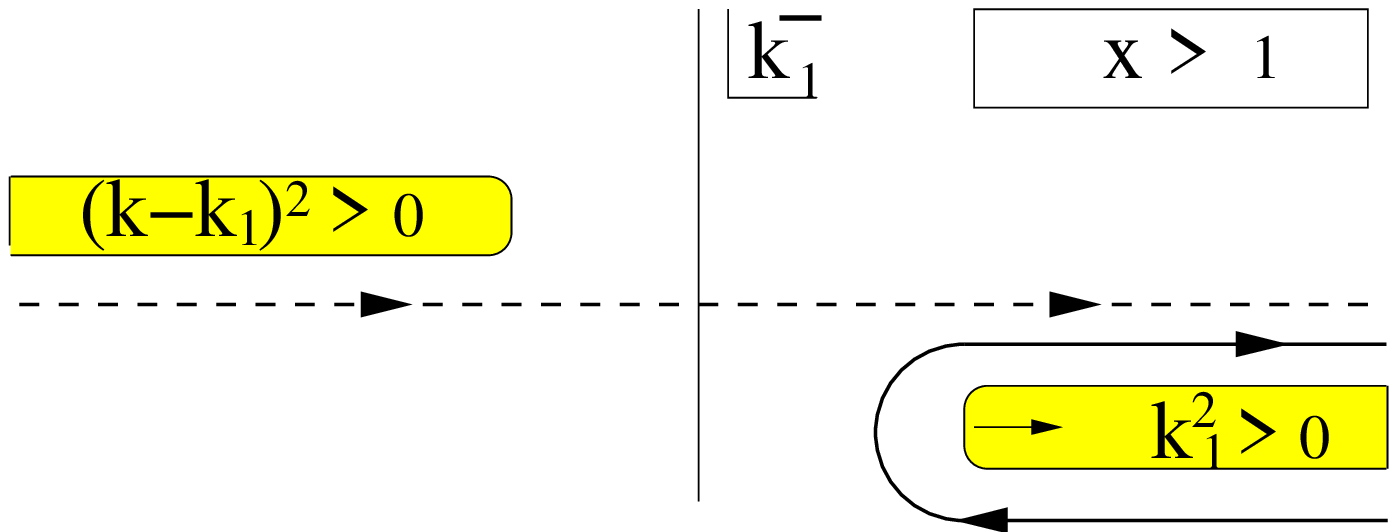}
(b) $\Delta_G(x,x)$
%\hspace{1.7cm}\mbox{}
\caption{\label{singularity-1} The integration contours
for the $k_1^-$ integration with respect to the singularities in
the amplitude ${\mathcal A}(k^2;s,u;s_1,u_1;k_1^2,(k-k_1)^2)$ 
relevant for gluonic pole contributions. The figure shows for
a given value of positive $s$ (relevant for $x>0$) how the $k_1^-$
integration bypasses the cuts in $s_1$, $u_1$ and the parton
virtualities in the limit $x_1\rightarrow +0$. The cases $0 < x < 1$
(distributions) and $x> 1$ (fragmentation) are shown in Figs (a)
and (b) respectively. In the latter case only parton virtualities
are relevant.
}
\end{figure}
 Again, parton or hadron masses as well as transverse
momenta have little bearing  as all they do is  move 
the endpoints of the cuts.  However, it is important to note 
that   $k_{1T}$ in the complete expression for the 
numerator of Eq.~(\ref{virts}) 
protects against the cut starting at zero in the zero mass limit. 
Depending on the value of $x_1$, 
the integration contour in $k_{1}^{-}$ bypasses the
singularities encountered in the complex plane in a particular
way, which dictates
the support properties of the quark-gluon-quark correlation functions.
The denominators in Eq.~(\ref{virts}) in the expressions relating $k_1^-$
to $s_1$ and $u_1$ tell us that only when $x_1 \in [x-1,1]$ (for positive
$x$) or $x_1 \in [-1,x+1]$ (for negative $x$) the singularities in $s_1$
and $u_1$ are relevant. 
We  study the case of the $s$-channel ($x > 0$). The $u$-channel is
analogous. 
Looking at gluonic poles, we  consider the limit $x_1 \rightarrow 0$.
For $0 < x < 1$, the value $x_1 = 0$ lies in the interval for which
the $s_1$ and $u_1$ discontinuities can contribute. 
These are shown in Fig.~\ref{singularity-1}(a), 
now together with the singularities
arising from the parton virtualities $k_1^2$ and $(k-k_1)^2$.
For the case $x > 1$ only these parton virtualities matter, shown
in Fig.~\ref{singularity-1}(b). 
We then find (including for $\Phi_G$ also the $u$-channel contribution)
in the limit $x_1\rightarrow 0$
\bea
\Phi_G(x,x) &=& 
\theta(x)\,\theta(1-x)\,\mbox{Disc}_{[s,s_1]}\mathcal{A}
\nn && + \theta(-x)\,\theta(1+x)\, \mbox{Disc}_{[u,u_1]}\mathcal{A},
\label{result-1}
\\
\Delta_G(x,x) &=& 0,
\label{result-2}
\eea
where for $\Delta_G(x,x)$ the $k_1^-$ integration can be wrapped
around the $k_1^2$ cut, which smoothly vanishes for $x_1 \rightarrow +0$.
This is described by the arrow inside the branch cut 
in  Figs.~\ref{singularity-1}(a) and (b), indicating that it
harmlessly recedes to infinity. Moreover it matches
continuously to the case that $x_1 < 0$.
Starting from $x_1 \rightarrow -0$ one immediately would have obtained
Eqs.~(\ref{result-1}) and (\ref{result-2}), since the $k_1^2$ cut is then
along the negative $k_1^-$ axis. This establishes the proof that 
gluonic pole matrix elements for fragmentation correlators vanish.
Similar results were obtained in the spectator model 
field theory in Ref.~\cite{Collins:2004nx}, the spectator approach 
in Ref.~\cite{Gamberg:2008yt}, and the general spectral approach of 
Ref.~\cite{Meissner:2008yf}.  But now it has been demonstrated in a 
completely general way 
by assuming unitarity and analyticity properties to hold for QCD. 
Similar to the earlier discussion,  
we note that the integration of Eq.~(\ref{GP})
and ${\mathcal A}(k^2;s,u;s_1,u_1;k_1^2,(k-k_1)^2)$ over $k_{1\sT}$
is ultraviolet divergent~\cite{Jaffe:1983hp,Diehl:1998sm,Bacchetta:2008xw}. 
 Again,  the  integral over $k_1^-$ alone is well behaved when $k_1^+$ 
and $k_{1\sT}$ are fixed with again full integration over $k_{1\sT}$ 
corresponding to appropriate  regularisation  of 
${\mathcal A}(k^2;s,u;s_1,u_1;k_1^2,(k-k_1)^2)$~\cite{Kang:2010xv}  
and study of the QCD evolution~\cite{Aybat:2011zv}.

 In our last step, we show that our arguments for vanishing
gluonic pole matrix elements hold for general multi-gluonic
and even multi-partonic pole matrix elements. Considering 
the analytic properties of general multi-gluonic pole matrix elements  
we can proceed inductively. For two gluons one simply extends the 
nesting of momenta $k-k_1$ and $k_1$ by a nesting
$k-k_1-k_2$, $k_1-k_2$ and $k_2$, which adds to the 
set ($s$, $u$, $s_1$, $u_1$) two new invariants ($s_2$, $u_2$), 
without changing the behavior in the others. The gluonic
pole matrix element $\Delta_{GG}(x,x,x)$ thus disappears as do
all higher pole matrix elements. Since these higher pole
matrix elements appear in the higher $k_\sT$-moments of the
correlator $\Delta^{[\mathcal U]}_{ij}(z,k_\sT)$ 
in Eq.~(\ref{TMDFF}), we conclude based on our very general
assumptions of analyticity for QCD amplitudes
that this TMD correlator is universal and will be
convoluted with the standard partonic cross sections.
There is no proliferation of functions originating from the structure
of the gauge links~\cite{Bomhof:2006ra}. 
This universality thus applies to all TMD fragmentation functions, 
T-even or T-odd, and for quark as well as for gluon PFFs.
Most well-known are the T-odd ones for quarks such as the Collins function 
$H_1^\perp$~\cite{Collins:1992kk,Boer:1997nt} and the polarization 
fragmentation function $D^\perp_{1T}$~\cite{Boer:1997nt,Boer:2010ya}.
These functions are simply allowed T-odd parts 
in the fragmentation correlator $\Delta(z,k_\sT)$ being a decay function.
The corresponding T-odd TMD distribution functions, the
Boer-Mulders function $h_1^\perp$ and the Sivers function $f_{1T}^\perp$ 
originate from the difference $\Phi^{[\mathcal U_+]}(x,p_\sT) 
-\Phi^{[\mathcal U_-]}(x,p_\sT)$ 
of correlators with different gauge links and as a consequence
will be convoluted with non-standard gluonic pole cross 
sections~\cite{Bacchetta:2005rm,Bomhof:2006ra,Bomhof:2007xt}. 
Our result also implies universality for the TMD fragmentation functions
of gluons~\cite{Mulders:2000sh}; including for instance the T-even 
TMD fragmentation functions $H_1^{\perp (g)}$, which just as the
corresponding distribution function $h_1^{\perp (g)}$ has a non-trivial 
gauge link dependence~\cite{Bomhof:2007xt,Dominguez:2011wm,Boer:2010zf}.
The T-even fragmentation functions $H_1^{\perp (g)}$, however, is
universal. In the case of these T-even functions, the non-trivial 
gauge link dependence only becomes visible in even $k_\sT$-moments 
involving contributions from T-even multi-gluonic 
pole matrix elements with an even number of gluons, all of which for 
fragmentation functions, however, will vanish.

\vspace{0.75cm}

\begin{acknowledgments}
\vskip -.25cm
This research is part of the Integrated Infrastructure
Initiative Hadron Physics 2 (Grant 227431).
LG acknowledges support from  U.S. Department of Energy under contract
DE-FG02-07ER41460. 
AM thanks BRNS (sanction no. 2007/37/60/BRNS/2913) for support.
\end{acknowledgments}
%\vspace{-1cm}

\bibliographystyle{h-physrev}

\bibliography{references}

\end{document}